\documentclass[twocolumn]{aastex63} 
\usepackage[utf8]{inputenc}
\usepackage{amsmath}
\usepackage{xcolor}
\usepackage{appendix}
\usepackage{longtable}
\usepackage{tablefootnote}
\usepackage{comment}
\usepackage{url}
\usepackage{enumitem}
\setlist[itemize]{noitemsep}

\newcolumntype{?}{!{\vrule width 4pt}}

\begin{document}

\title{What's the (RV) Point? A $3.5\times$ Enhancement in Super-Jupiters with Saturn-like Periods from a Critical Observation}
\author[0000-0002-6832-4680]{Marie C. Tagliavia}
\affiliation{Department of Physics and Astronomy, University of Notre Dame, Notre Dame, IN 46556, USA}
\author[0000-0002-3725-3058]{Lauren M. Weiss}
\affiliation{Department of Physics and Astronomy, University of Notre Dame, Notre Dame, IN 46556, USA}

\begin{abstract}
Amidst the exoplanet revolution in which multiple techniques have successfully found planets, the Doppler (Radial Velocity, or ``RV'') technique is unique in its sensitivity to giant planets at long orbital periods around Sun-like stars.  The upcoming retirement of Keck-HIRES will incur irreversible changes in the continuation of HIRES's decades-long stable RV baseline and with it, the exoplanet community's ability to detect giant exoplanets with periods longer than Jupiter. With the time elapsed from the last HIRES RV for many stars of interest at $\sim3$ years and growing, we tested the impact of a ``critical RV", one that would bridge this gap between past HIRES RVs and future stable Keck-KPF RVs, on the recovery of long-period giant exoplanets. We generated 2000 1-planet systems with RVs sampled at a representative timeseries and used the planet-finding code \texttt{Octofitter} to perform injection-recovery experiments including and omitting this critical RV for each system. For the injected long-period super-Jupiters ($\sim 8-55$ years, $1-13 M_J$), including the critical RV induced a $1.5\times$ enhancement in overall planet recovery and a more specific $3.5\times$ enhancement in the recovery of super-Jupiters with Saturn-like periods. These experiments show that gathering a critical RV for stars of interest can help ensure that HIRES's decades-long stable RV baseline in conjunction with future KPF RVs, or indeed that the observationally-gapped RV baselines of any instruments that will undergo an RV zeropoint offset, will continue to be foundational to the discovery of long-period giant exoplanets in years to come.
\end{abstract}

\section{Introduction}
Long-period giant exoplanets are of great value in understanding the planetary systems they inhabit. As perhaps the most massive objects in their system besides the host star, these exo- and super-Jupiters can significantly impact the evolution and dynamics of smaller bodies, such as exo-Earths \citep{Kane2020,Kane2024}. There have not yet been definitive conclusions regarding the link between cold Jupiters and inner rocky exoplanets --- theoretical work suggests alternatively positive \citep{ChiangLaughlin2013} and negative correlations \citep{Izidoro2015, Lambrechts2019}.  Observational and statistical studies also have not yet converged; even after decades of RV observation \citep{Weiss2024,Bonomo2023}, some studies find a modest positive correlation \citep{Bryan2024, Zhu2024, VanZandt2025}, some no correlation at all \citep{Bonomo2025}, and some consistent with a modest negative correlation \citep{Bonomo2023}. Therefore, discovering more of these long-period giants and characterizing their planetary systems is vital for refining theoretical work that posits the criteria for habitability, especially as the era of the NASA Habitable Worlds Observatory (HWO) approaches.

Demographics studies based on RV data \citep{Fernandes2019, Fulton2021} as well as direct imaging data \citep{Nielsen2019} point to a peak in the giant planet occurrence rate of $\sim15\%$ at a semimajor axis of $\sim 3$ to $10$ AU (orbital period $\sim 5$ to $32$ years for Sun-like stars), with a possible falloff in planet occurrence to $\sim9\%$ at larger separations (10 to 30 AU), while recent work has constrained this peak to $2-8$ AU for $1-13 M_J$ planets around $1.2M_\odot$ and higher (Nielsen et al. 2026, in prep.).  The large orbital separations associated with the giant planet peak and occurrence falloff are a region of parameter space that growing RV baselines have just recently been able to access in depth. 

The discovery and orbital characterization of long-period giant exoplanets requires long-term, dedicated, time-domain surveys that at least approach, and ideally exceed, the timescales of the orbits of the planets. The W. M. Keck Observatory's HIgh Resolution Echelle Spectrograph (HIRES, \citealt{Vogt1994}) instrument has been at the forefront of such long-term surveys.  With HIRES, potential planet-hosting stars are monitored for decades via the radial velocity (RV) technique in the hope of detecting the gravitational reflex of the star due to the presence of one or more long-period planets (e.g., \citealt{Rosenthal2021} and references therein).

Inherent in HIRES's success is that it has been stable for decades, with only one major instrument change: a CCD upgrade in 2004 \citep{McLean2006}.  Meanwhile, the Keck Planet Finder (KPF, \citealt{Gibson2024}) is a new instrument at Keck that has achieved exceptional RV precision on short timescales---those spanning a single night to two weeks (e.g., \citealt{Hon2024,Rubenzahl2023}).  Despite these early accomplishments, KPF has not yet demonstrated a stable baseline on a timescale of $\sim1$ year, which is a prerequisite for sensitivity to planets on long-period orbits.  Such a baseline is unlikely to be applied retroactively; since its first light in November 2022, KPF has undergone multiple servicing missions, each of which incurred an RV zeropoint offset.  It is unclear whether these early epochs of observation can be cleanly stitched into a long-term RV time series, particularly for systems that were observed at low to moderate cadence.  As a result, the spectra of many systems of interest, including 63 of the HWO Tier 1 targets that are observable from the northern hemisphere, are not yet contributing to the long-term RV monitoring that is necessary for planet discovery at long time-scales.

The three-year (and increasing) data gap between the last HIRES RVs and the first stable KPF RVs for these targets is a cause for concern.  How many planets (and what types of planets) are we missing due to such a gap in the data?  Here, we explore the role of a single, time-critical RV, to be taken with HIRES, that would close the widening gap between HIRES and KPF. While our study is motivated by the specific instrument migration from HIRES to KPF at WMKO, these results are likely applicable to instrument migrations, upgrades, or other actions that would induce an RV zeropoint offset at other facilities.

\section{Methods}
To conduct a controlled experiment, we generated synthetic planetary systems, for which the underlying architecture is known.  We then tested how these systems, sampled with a realistic time series and RV errors, would fare when subjected to a commonly-used planet finding package, both with and without a ``critical'' RV.  This critical RV is defined as a continuation of the stable RV baseline from a single instrument across an observational gap (Fig. \ref{fig:timeseries_rvs}). We hypothesized that the inclusion of this critical RV would allow for a better solution of the RV zeropoint offset between our nominal HIRES and KPF instruments and thus lead to the recovery of more planets.

\subsection{Generating Synthetic Systems \label{sec:gen}}
We adopted the time series of a long-observed HIRES target, HD 213472, as the basis of our study.  The time series has 13 instances of \texttt{hires\_k} measurements, followed by 64 instances of \texttt{hires\_j} measurements.  The overall number of RVs is fairly typical of the 719 stars observed as part of the California-Legacy Survey on HIRES \citep{Rosenthal2021}. The reason we chose this time series is that it is well-suited to the question we are trying to test: the data fit our criteria for a critical RV (see Fig. \ref{fig:timeseries_rvs}). While the data are originally from HIRES both before (\texttt{hires\_k}) and after (\texttt{hires\_j}) an instrument upgrade, we modified the data at these time-stamps in a manner suitable for studying the HIRES-to-KPF transition, as described in the remainder of this section.

\subsubsection{Timestamps}
\begin{figure*}
    \centering
    \includegraphics[width=\textwidth]{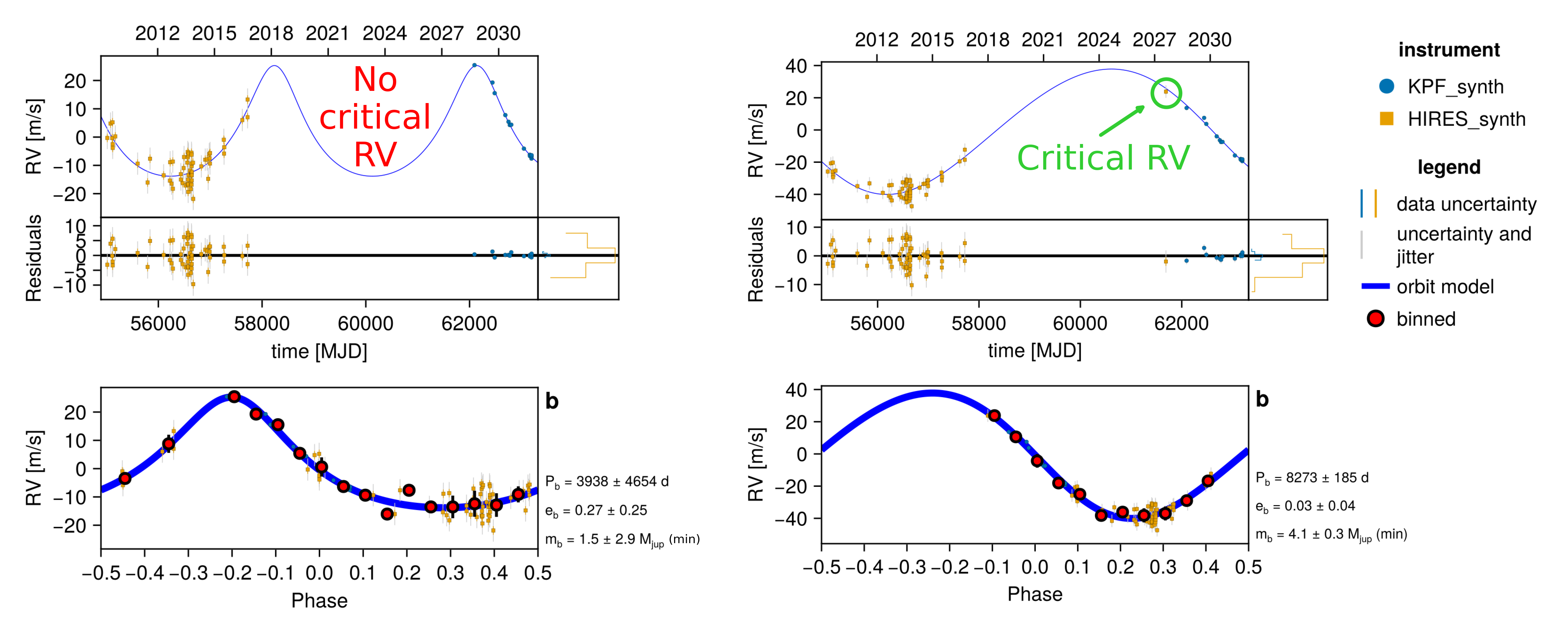}\hfill
    \caption{The outcomes of the \texttt{Octofitter} runs without (left) and with (right) the critical RV on synthetic system 23, whose planet has $P=8113.4$ days and $M_\mathrm{pl}=4.46 M_J$ and which was selected as a clear demonstration of the effect of including a critical RV on accurate and precise planet recovery. In each plot, produced by \texttt{CairoMakie} within Octofitter, the \texttt{HIRES\_synth} data (orange) appear first in the timeseries, followed by an observational gap, after which the critical RV (within the \texttt{HIRES\_synth} data, circled in green) was taken immediately preceding the collection of the \texttt{KPF\_synth} data (blue). The values seen to the right of the lower subplots are the means and standard deviations of the posteriors for each run. Note that, although the goodness-of-fits are nearly equivalent, the run \textit{with} the critical RV recovers much more accurate and precise parameter values (see Appendix \ref{appendix:A} for more detail).}
    \label{fig:timeseries_rvs}
\end{figure*}

To mimic how the gap induced by the transition from HIRES to KPF affects planet sensitivity, we (1) reversed the time series (note that this is entirely aesthetic), and (2) assigned the timestamps from \texttt{hires\_k} as presumed future KPF observation times (Figure \ref{fig:timeseries_rvs}).  Also note that we only used the timestamps of these observations; our handling of the RVs and RV errors is discussed below.

\subsubsection{RVs and RV errors}
We generated 2000 simulated systems, sampled at our timeseries, to produce RVs. Each system had a star of 1 solar mass orbited by 1 planet, on an edge-on, circular orbit, with a mass ($M_\mathrm{pl}$)\footnote{Equivalently, one could interpret our experiment as varying $M_\mathrm{pl}$sin$i$ and being agnostic to $i$.} drawn from a log-uniform distribution of $1-13 M_J$ and an orbital period ($P$) drawn from a log-uniform distribution of $3000 - 20000$ days (i.e. $8.21-54.76$ years). We applied a phase-shift ($\tau$) to the orbit at the time of first RV measurement drawn uniformly from [0,1). Systems with eccentric orbits and/or multiple planets would be of interest, but are outside the scope of this work (see Section~\ref{subsec:future_directions}).

We then added errors to the RVs. For the first part of the timeseries (representative of HIRES data and denoted \texttt{\texttt{HIRES\_synth}} hereafter), we added a Gaussian-distributed RV error of 1.5 m/s to represent the intrinsic measurement errors ($\sigma_\mathrm{HIRES, int}$), and an additional Gaussian-distributed RV error of 3 m/s to represent the RV jitter ($\sigma_\mathrm{HIRES, jit}$).  For the second part of the timeseries (representative of the KPF data and denoted \texttt{\texttt{KPF\_synth}} hereafter), we added a Gaussian-distributed error of 0.5 m/s to represent the intrinsic instrument error ($\sigma_\mathrm{KPF, int}$), and an additional Gaussian-distributed error of 2 m/s to represent the RV jitter ($\sigma_\mathrm{KPF, jit}$).  Note that KPF is assumed to have superior per-measurement RV precision to HIRES, and datasets from both instruments are assumed to have RV jitter that is a stand-in for the combination of stellar activity and instrument systematics.  No other types of error (e.g., correlated RV errors from stellar activity) were added to the synthetic data.  Finally, we introduced an RV zeropoint offset for each RV dataset ($\gamma_\mathrm{HIRES}$, $\gamma_\mathrm{KPF}$), drawing each value from a uniform distribution spanning $[-150,150]$\,m/s.

Altogether, the following formula was used to generate the RVs for the simulated systems:
\begin{equation}
\begin{split}
    \hspace{-0.5cm}
    \mathrm{RV}(t,\mathrm{inst}) &= 28.4 \mathrm{m/s} \bigg( \frac{P}{\mathrm{yr}} \bigg)^{-\frac{1}{3}} \bigg( \frac{M_{pl}}{M_J} \bigg) \cos{\Bigg( 2\pi \bigg(\frac{(t - t_0)}{P} - \tau\bigg) \Bigg)} \\
    &\quad + \mathrm{Normal}(0,\sigma_{\mathrm{inst, int}}) + \mathrm{Normal}(0,\sigma_{\mathrm{inst,jit}}) + \gamma_\mathrm{inst} \\
\end{split}
\end{equation}
where $t_0$ is the time at first RV collection and $\sigma_{\mathrm{inst, int}}$, $\sigma_{\mathrm{inst,jit}}$, and $\gamma_\mathrm{inst}$ represent the appropriate instrument variables for a certain point in the selected timeseries for a given instrument. A summary of the distributions from which we drew parameters for generation can be found in Table \ref{tab:params}.

\begin{deluxetable}{ll}
\tablecaption{\label{tab:params} Synthetic RV Generation}
\tablehead{
\colhead{Parameter} &
\colhead{Distribution}
}
\startdata
{$P$ (years)} & LogUniform(8.21, 54.76) \\
{$M_\mathrm{pl}$ ($M_J$)} & LogUniform(1,13) \\ 
{$\tau$} & Uniform(0,1) \\ 
{$\sigma_\mathrm{HIRES, int}$ (m/s)} & 1.5 \\
{$\sigma_\mathrm{HIRES, jit}$ (m/s)} & 3 \\
{$\gamma_\mathrm{HIRES}$ (m/s)} & Uniform(-150,150) \\
{$\sigma_\mathrm{KPF, int}$ (m/s)} & 0.5 \\
{$\sigma_\mathrm{KPF, jit}$ (m/s)} & 2 \\ 
{$\gamma_\mathrm{KPF}$ (m/s)} & Uniform(-150,150) \\ 
\enddata
\end{deluxetable}

\subsection{Planet Recovery}
\label{subsec:rec_criteria}
For each system, we used the publicly-available planet-finding package \texttt{Octofitter} to attempt to recover the injected planet \citep{Thompson_2023}. \texttt{Octofitter} uses \texttt{Pigeons}, which is a parallel-tempered, no-U-turn implementation of Markov-Chain Monte Carlo (MCMC) that reliably samples from multi-modal posteriors \citep{pigeons, surjanovic2023pigeons}.  We explored two cases: (1) with the critical RV datapoint, and (2) without it.  

\subsubsection{The Model}
For simplicity, we assumed there was exactly one planet in the RV time series.  A search that allows for multiple planets (or no planets) would provide a more realistic assessment but is outside the scope of this work.  We allowed the one-planet fit to have eccentricity, resulting in recovered parameters for both the planet eccentricity and longitude of periastron (even though all injected planet orbits were circular).  We also fit an RV jitter and RV zeropoint offset for each instrument.  Note that \texttt{Octofitter} requires a constant \texttt{t\_ref}; we set this equal to the time at first RV observation.

Thus, we fit 9 free parameters in total: the orbital period $P$, planet mass $M_\mathrm{pl}$, orbit fraction completed with respect to periastron at first measurement $\tau$, eccentricity $e$, longitude of periastron $\omega$, \texttt{HIRES\_synth} RV jitter $\sigma_\mathrm{HIRES, jit}$, \texttt{HIRES\_synth} RV offset $\gamma_\mathrm{HIRES}$, \texttt{KPF\_synth} RV jitter $\sigma_\mathrm{KPF, jit}$, and \texttt{KPF\_synth} RV offset $\gamma_\mathrm{KPF}$.

\subsubsection{Priors}
To discourage selection of periods significantly longer than the RV baseline ($\approx$22 years) during sampling, we adopted the baseline prior as previously used in the transit \citep{Vanderburg2016, Kipping2018} and RV \citep{Blunt2019} communities. This prior is uniform on $(P_{min}, T_{obs})$ and log-uniform on $(T_{obs}, P_{max})$, where $T_{obs}$ is the RV baseline. We impose a beta prior on eccentricity, with optimal values of $a=0.867$ and $b=3.03$ adopted from \cite{Kipping2013}. This distribution was then truncated to $[0.001,0.999]$ to prevent selection of nonphysical values of eccentricity and to prevent code failure at extrema. The priors for all parameters are detailed in Table \ref{tab:priors}. We then ran \texttt{Octofitter} using the \texttt{Pigeons} optimization algorithm, yielding posterior distributions of parameters for each synthetic planetary system.

\begin{deluxetable}{ll}
\tablecaption{Priors used in Octofitter recovery\label{tab:priors}}
\tablehead{
\colhead{Parameter} &
\colhead{Distribution}
}
\startdata
{$P$ (years)} & Baseline($10^{-3}$,$10^5$,$T_{obs}$) \\
{$M_\mathrm{pl}$ ($M_J$)} & Uniform(0,1000) \\ 
{$\tau$} & UniformCircular(0,1) \\
{$e$} & TruncatedBeta(0.867,3.03) \\ 
{} & on [0.001,0.999] \\ 
{$\omega$} & UniformCircular(0,$2\pi$) \\ 
{$\sigma_\mathrm{HIRES, jit}$ (m/s)} & Uniform(0,10) \\
{$\gamma_\mathrm{HIRES}$ (m/s)} & Normal(0,150) \\
{$\sigma_\mathrm{KPF, jit}$ (m/s)} & Uniform(0,10) \\ 
{$\gamma_\mathrm{KPF}$ (m/s)} & Normal(0,150) \\ 
\enddata
\end{deluxetable}

\subsubsection{Recovery Classification}
For any injected signal to be classified as recovered (i.e. detected), it had to meet both precision and accuracy criteria. We required that the posteriors have sufficiently narrow distributions centered around the following injected values: $\sigma_P/P < 0.3$, $\sigma_{M_\mathrm{pl}\sin{i}}/M_\mathrm{pl}\sin{i} < 0.2$, and $e < 0.1$. We also required the recovered period and mass to be within 20\% of the injected values. These criteria were chosen somewhat arbitrarily --- either looser (or tighter) criteria could be applied, resulting in more (or fewer) recoveries.   Note also that our recovery algorithm always adopts a one-planet model,  whereas other schemes first compare an N planet model to an N+1 planet model via a statistical metric (e.g., \citealt{Rosenthal2021}, which uses a difference in Bayesian Information Criterion for model selection) and then apply similar precision and/or accuracy criteria to the injected vs. recovered signal for dispositioning which signals count as ``recovered.''  Nonetheless, in our scheme, the same definition of recovery is applied to the datasets with and without the critical RV, allowing a robust comparison.

\section{Results}
\label{sec:results}
\begin{figure*}
    \centering
    \includegraphics[width=\textwidth]{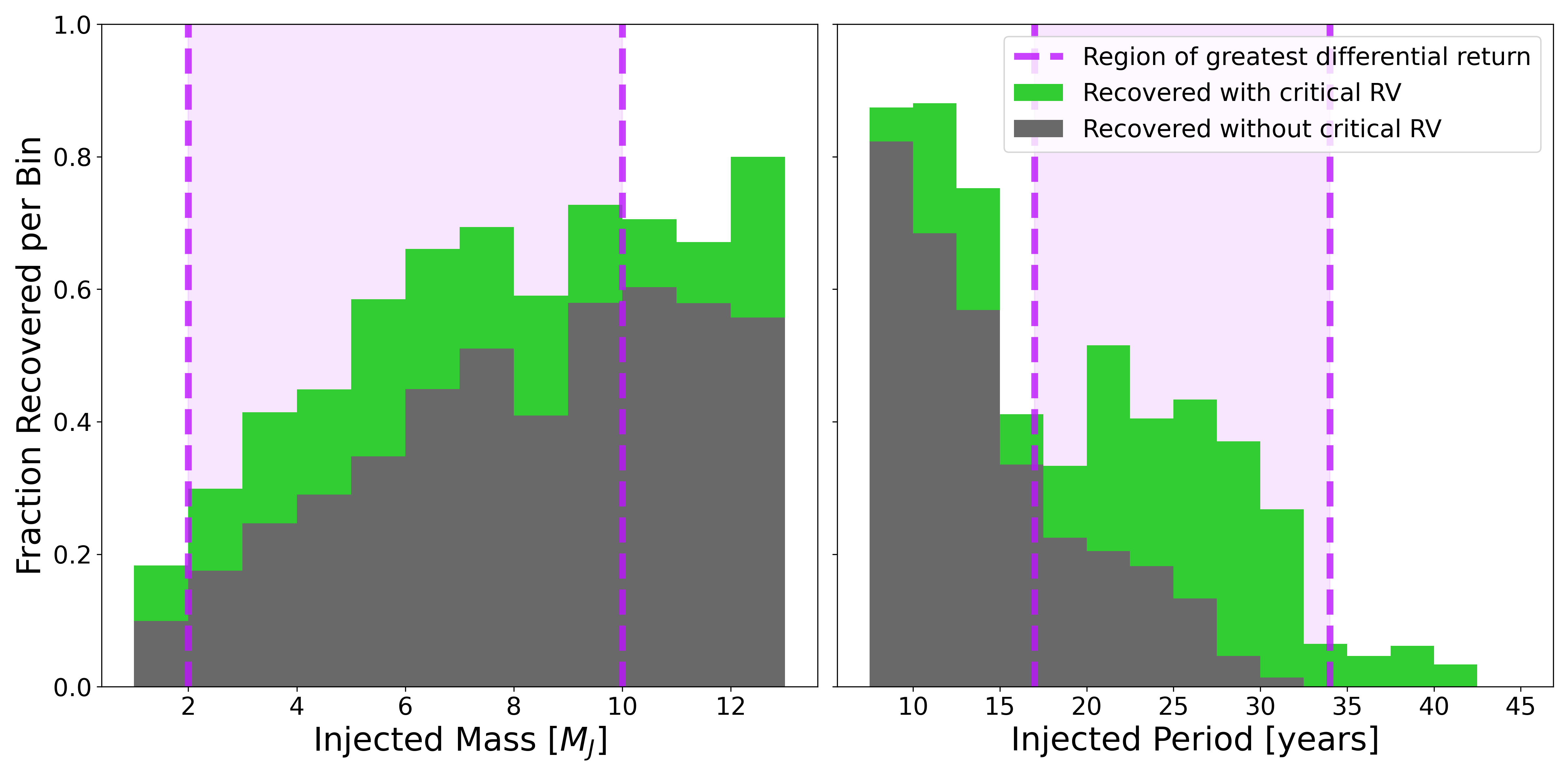}\hfill
    \caption{
    Histograms of the proportion of planets recovered with (green) vs. without (gray) the critical RV, normalized to the number of planets injected in each bin. Note that, rather than being stacked, both histograms start from zero, such that the visible portions of the green histogram correspond exactly to the ``recovered only with critical RV'' points in Figure \ref{fig:recovered_planets}.}
    \label{fig:histograms}
\end{figure*}

\begin{figure*}
    \centering
    \includegraphics[width=\textwidth]{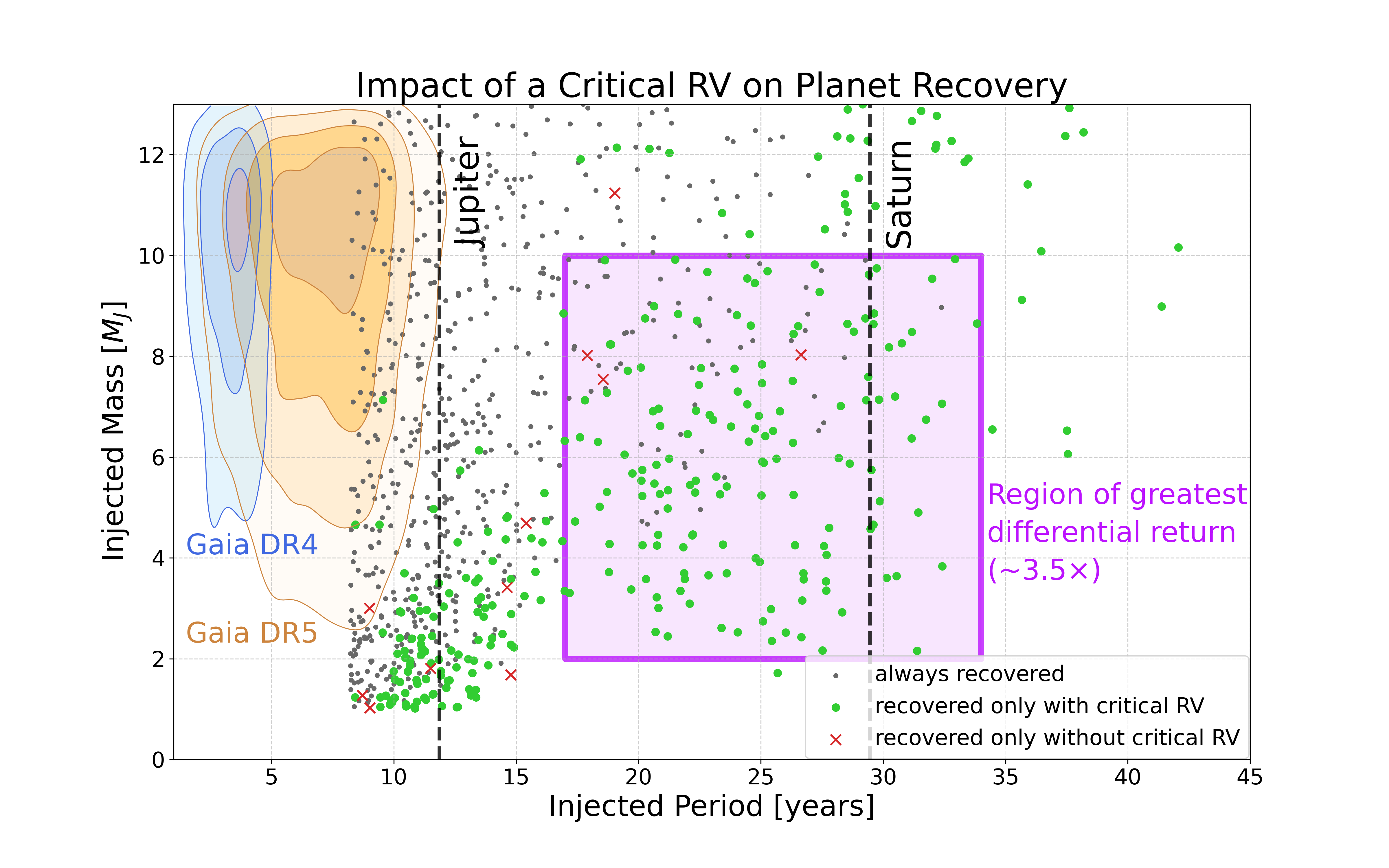}\hfill
    \caption{The injected mass vs. orbital period of planets that were recovered: (1) both with and without the critical RV (gray circles), (2) only with the critical RV (green circles), and (3) only without the critical RV (red x's).  The expected planetary yield from Gaia DR4 and DR5 astrometry (blue and orange contours, \citealt{Lammers_2026}) and the orbital periods of Jupiter and Saturn (dashed lines) are overlaid for context. The purple box highlights the region of greatest differential return.}
    \label{fig:recovered_planets}
\end{figure*}

Table \ref{tab:inj_rec_values} in Appendix \ref{appendix:A} presents the detailed results of our experiment.  Each row includes the injected values for one planetary system along with its recovered properties from the Octofitter runs with and without the critical RV included. These results are also represented in Figures \ref{fig:histograms} and \ref{fig:recovered_planets}.  

Figure \ref{fig:histograms} displays the proportion of planets recovered with vs. without the critical RV, normalized to the number of planets injected in each bin. Many of the injections that are only recovered with the critical RV are (1) at long periods, or (2) at low planet masses. As expected, there are more planets recovered with than without the critical RV in each bin, and the recovery proportion increases nearly monotonically with mass. Many planets are recovered with the critical RV, but missed without it, at $\sim 25$ years. This feature informed our determination of the ``region of greatest differential return'' for planet detections (denoted in purple in Figures \ref{fig:histograms} and \ref{fig:recovered_planets}). This especially-notable region of parameter space from 17-34 years and 2-10 Jupiter masses is taken to represent a population of possible super-Jupiters with Saturn-like periods that we stand to miss if we do not collect the critical RV.  Note that while the exact boundaries are chosen to highlight this significant increase in planet recovery, this region is broad, spanning a factor of 2 in orbital period and a factor of 5 in mass, and moderate redefinitions of these boundaries lead to similar enhancements.

To compute the enhancement of detected planets due to the critical RV, we consider the number of injections, number of true recoveries, and number of false recoveries.  We injected 2000 planets over the entire parameter space, and for each planet, recovery was attempted with, then without, the critical RV.  Overall, 568 planets were recovered both with and without the critical RV ($N_\mathrm{both}$), whereas 296 planets were recovered only using the data set including the critical RV and were missed otherwise ($N_\mathrm{flip}$).  There were also 11 systems for which the planet was only recovered using the data set without the critical RV (but not with it) --- these likely represent our error ($N_\mathrm{err}$).  Computing the fractional enhancement in planets detected due to the inclusion of the critical RV as:
\begin{equation}
    Q = \frac{N_\mathrm{both} + N_\mathrm{flip} - N_\mathrm{err}}{N_\mathrm{both}}
\end{equation}
yields a $1.5\times$ enhancement in detected planets, corresponding to a $50.18\%$ increase in detections. Within the region of greatest differential return, we injected 487 planets, while recovery resulted in values of $N_\mathrm{both} = 57$ planets, $N_\mathrm{flip} = 146$ planets, and $N_\mathrm{err} = 3$ planets. Thus, the critical RV enhanced the number of detected super-Jupiters with Saturn-like periods by a factor of $3.5\times$, corresponding to a $250\%$ increase in detections.

\section{Discussion}
We found that for a representative time series, there can be a critical RV datapoint acquired after a long observational hiatus, and immediately preceding the regular operations of a new RV instrument.  This critical RV plays an out-sized role in planet detection, especially for long-period giant planets, resulting in a $3.5\times$ enhancement in the number of super-Jupiters with Saturn-like periods detected.

\subsection{Possible Experiment Permutations}
\label{subsec:future_directions}
Our experiment was restricted to a single time series and a very simple planet model.  In principle, many permutations to this experiment are possible.  We can qualitatively predict how such tests might alter the fractional enhancement $Q$ and/or the region of greatest differential return.  For instance, the following modifications would likely alter the region of greatest differential return:
\begin{itemize}
    \item Lengthening or reducing the gap in RV coverage preceding the critical RV,
    \item Lengthening or reducing the span of \texttt{\texttt{KPF\_synth}} RVs,
    \item Lengthening or reducing the total RV baseline.
\end{itemize}
Here, lengthening data sets and/or the gap would probably increase the orbital periods associated with the region of greatest differential return.

Adding a second critical RV (i.e., another RV observation from the first instrument, after the data gap) or shifting the critical (\texttt{HIRES\_synth}) RV within the span of \texttt{KPF\_synth} RVs to create a stronger overlap between the instruments might yield an even higher value of $Q$ than what we found here.  Experiments that relax our single, circular-planet model might yield lower values of $Q$:
\begin{itemize}
    \item Injecting planets with non-zero eccentricity,
    \item Injecting multiple planets,
    \item Running fits for numbers of planets other than the number of planets injected.
\end{itemize}

These experiments might be interesting, but would be time-consuming to pursue in their entirety. Due to our timeline being motivated by current instrument events, we leave them to future work.

\subsection{Relevance in the Gaia era}

The Gaia mission represents an enormous advancement in the detection of exoplanets via astrometry. At the time of writing, just three confirmed planets (with $M_p < 13 M_J$) have been initially detected via astrometry \citep{Curiel2022, Sozzetti2023, Stefansson2025}. However, the releases of Gaia DR4 and DR5 are expected to bring about the discovery of $1900 \pm 540$ and $38000 \pm 7300$ planets, respectively, with well-determined orbital periods and masses; these are predicted to be primarily super-Jupiters with periods from $\sim 1-12$ years (\citealt{Lammers_2026}, see contours in Figure \ref{fig:recovered_planets}).  Gaia astrometry will also enable updated orbit fits and the determination of true planet masses, instead of minimum masses, for planets that were previously discovered with RVs \citep[e.g.][]{VanZandt2026}.

Although this orders-of-magnitude increase will eventually greatly enhance the population of exoplanets, these discoveries are still at least a year away for DR4 and several years away for DR5. Moreover, this region of greatest sensitivity for Gaia is a factor of $\sim 3-5$ shorter in period than the parameter space of greatest differential return from the addition of a critical RV (see Figure \ref{fig:recovered_planets}).  Unlike RV surveys, Gaia is limited in its time baseline because the mission flew for 10.5 years, whereas RV surveys can be extended and, given appropriate resources, gracefully transitioned from one instrument to another.  Thus, the RV discovery method provides a crucial advantage for discovering super-Jupiters with Saturn-like periods.

\subsection{Relevance in the WMKO Ecosystem}
The main motivation in conducting this study was to characterize the scientific impact of an early HIRES retirement.  This is relevant because there is pressure in two directions: within the WMKO ecosystem, there is a need to free space on the Nasmyth deck (priv. comm.), whereas from the exoplanet community perspective, there is a need for HIRES to persist long enough to get a critical RV for systems that need it.

The systems for which a critical RV is most relevant are those with long-term RV monitoring on HIRES, which are also likely to be studied with KPF.  Example programs are the Habitable Worlds Observatory precursor program (HWO; PI Weiss), the California Legacy Survey (CLS, \citealt{Rosenthal2021}), the TESS-Keck Survey (TKS, \citealt{Chontos2022}), and the Kepler Giant Planet Search (KGPS, \citealt{Weiss2024}).  The median time in days since the most recent HIRES RV for these programs is provided in Table \ref{tab:hires_programs}.  These times correspond to data gaps of $\sim3$ to 4 years, and are growing.  

For some targets, especially bright stars, additional facilities such as Lick Observatory's Automated Planet Finder/Levy Spectrograph \citep{Radovan2014} the Telescopio Nazionale Galileo's HARPS-N \citep{Cosentino}, the Discovery Channel Telescope's EXPRES \citep{Jurgenson2016}, and the WIYN Telescope's NEID \citep{Schwab2016, Halverson2016} might help bridge the gap between HIRES and KPF, but such observations are not guaranteed.  For faint stars ($m_V > 10$), it is unlikely that these facilities, which include smaller-aperture telescopes than Keck, will provide meaningful RV coverage.

The continued availability of HIRES, coupled with support from telescope time allocation committees, would allow a critical RV observation to be scheduled for each star in each of these programs.  Some of these programs are large --- in particular, CLS has 719 stars!  While the programs listed in Table \ref{tab:hires_programs} represent the bulk of long-term precision RV programs on HIRES for which critical RV collection may be desired, the RV community might have an additional vested interest in securing the critical RVs for historic HIRES targets that are not represented in these programs. 

\begin{deluxetable}{lrrr}
\tablecaption{HIRES RV Program Status\label{tab:hires_programs}}
\tablehead{
\colhead{RV Program} &
\colhead{N Stars} &
\colhead{$m_V$} &
\colhead{Time Since Last RV (days)$^\star$}
}
\startdata
HWO  & 67  & 5.3  & 1000 \\
CLS  & 719 & 7.6  & 1000 \\
TKS  & 144 & 10 & 1100 \\
KGPS & 63  & 12 & 1400 \\
\enddata
\tablecomments{Apparent visual magnitudes and times since last RV are median values for each program.  The RV programs are HWO = Habitable Worlds Observatory precursor program, CLS = California Legacy Survey \citep{Rosenthal2021}, TKS = TESS Keck Survey \citep{Chontos2022}, and KGPS = Kepler Giant Planet Search \citep{Weiss2024}.  $^\star$The time since last RV is determined from Jump, a database tool of  the \cite{CPSWebsite}.}
\end{deluxetable}

\subsection{A Nominal HIRES-KPF Bridge Program}
Based on Table \ref{tab:hires_programs}, we estimate that a minimal bridge program to migrate long-term HIRES programs to KPF would require 30 Keck nights.  Such an observing program would collect a critical RV with HIRES for each target, and also one KPF RV for each target --- although it is vital that the KPF RV is collected after KPF has demonstrated long-term stability. This program would require approximately 1000 new HIRES observations. Based on exposure times of historic HIRES observations available on Jump \citep{CPSWebsite}, if we approximate each visit (exposure + 2-minute overhead) as 7 minutes for each star in the CLS and HWO programs ($m_V \sim 7.5$, $m_V \sim 5$, respectively), 12 minutes for each star in TKS ($m_V\sim10$), and 22 minutes for each star in KGPS ($m_V\sim12$), the time to acquire the HIRES RVs is 8616 minutes, or approximately 15 Keck nights.  Assuming comparable exposure times with KPF yields a bridge program of 30 Keck nights.  These observations would need to be scheduled over a full year (or more) to account for the distribution of target right ascensions.  While 30 nights is an enormous ask within the Keck ecosystem (nearly 10\% of available Keck 1 time in a year, or 5\% if spread over 2 years) and would likely require the support of one or more major Keck partners, such an investment would enable the unique return of time-critical RVs that perpetuate the $\sim30$-year-long RV baselines of over 1000 stars and provide sensitivity to planets that are beyond the reach of other techniques.

\section{Conclusion}
Discovering long-period giant exoplanets is crucial not only to improving planetary demographics but also to better understanding the question of extrasolar habitability.  The discovery of such planets is a decades-long endeavor that spans the development of new instruments and the retirement of old ones.  In this work, we investigated what types of planets we stand to discover --- or miss --- as a function of observing strategy.  We found:
\begin{itemize}
    \item Collecting a critical RV that bridges a 10-year gap between data from two RV instruments leads to a $1.5\times$ enhancement in overall planet recoveries. 
    \item The critical RV leads to a $3.5\times$ enhancement in the recovery of super-Jupiters with Saturn-like periods ($2-10 M_J$, $17-34$ years).
    \item The parameter space where Gaia astrometry is expected to find planets is at much shorter orbital periods than the super-Jupiters to which the critical RV is especially sensitive.
    \item The region of greatest return of a critical RV probes both the observed peak in the giant planet occurrence rate and the falloff at larger orbital separations, underscoring the importance of long-term RV monitoring in the determination of the location of this peak.
    \item The 10-year gap adopted here is relevant to multiple historic RV surveys that have median data gaps of 3 years, with some targets having data gaps of 10 years, and some targets having been selected by the community as an optimal discovery space for habitable planets.
\end{itemize}

We hope that the results of this experiment prove useful in guiding policy decisions as observatories consider upgrades and/or replacements of their RV survey instruments.  We also hope these results will help the exoplanet community explain to telescope time allocation committees the potential scientific return of critical RVs.

\acknowledgments
M.C.T. and L.M.W. acknowledge support from the NASA Exoplanet Research Program (grant no. 80NSSC23K0269), a NASA-Keck PI data award (grant no. 80NSSC25K0188), and the University of Notre Dame's Arthur J. Schmitt Presidential Leadership Fellowship. This research was supported in part by the resources of the Notre Dame Center for Research Computing (CRC). M.C.T. particularly acknowledges the Astroweiss group for being a supportive intellectual community that offers much insight and encouragement, and also extends gratitude to Joshua N. Winn and his group for their welcoming attitude and productive discussions. Additionally, M.C.T. and L.M.W. thank Judah Van Zandt and Erik Petigura for improving this manuscript with their thoughtful comments on one of its earlier versions. This analysis made use of Jump, a database tool of the California Planet Search. A portion of the code used in this analysis was written with the assistance of ChatGPT (OpenAI). This analysis was based on archival data collected at the W. M. Keck Observatory at Maunakea.  We recognize and acknowledge the very significant cultural role and reverence that Maunakea has within the Native Hawaiian community. We are most fortunate to have the opportunity to analyze observations that were made from this mountain.

\facilities{None}

\software{Julia (v1.10.2), Octofitter (v7.0.0) \citep{Thompson_2023}, OctofitterRadialVelocity (v7.0.0) \citep{Thompson_2023}, Pigeons (v0.4.9) \citep{pigeons, surjanovic2023pigeons}, CairoMakie (v0.13.10) \citep{CairoMakie2021}. Python (v3.11.4), astropy (v5.3.2, v7.1.0) \citep{astropy:2013, astropy:2018, astropy:2022}, matplotlib (v3.7.1, v3.10.3) \citep{matplotlib2007}, numpy (v1.25.2, v2.3.0) \citep{numpy2020}, pandas (v2.0.0, v2.3.0) \citep{pandas2010, pandas2020}.}

\clearpage

\bibliography{main}{}
\bibliographystyle{aasjournal}

\appendix

\setcounter{table}{0}
\renewcommand{\thetable}{A\arabic{table}}

\section{Table of Injected and Recovered Values}
\label{appendix:A}

As discussed in Section \ref{sec:results} (Results), the tables within this appendix compile the results of our injection-recovery experiment. Each row of the data table comprises 1 system, where parameters ending in ``\texttt{\_inj}" refer to values injected and parameters ending in ``\texttt{\_fit\_wo}"/``\texttt{\_fit\_w}" refer to values recovered in the Octofitter run without/with the critical RV.

Included on the following page is Table \ref{tab:param_defs}, a description of every parameter included in the unabridged table within the CSV file. This is followed by Table \ref{tab:inj_rec_values}, a sample of the table of injected and recovered values. While the corresponding CSV includes all systems and all parameters detailed in Table \ref{tab:param_defs}, the abridged Table \ref{tab:inj_rec_values} includes only select systems as well as only those parameters most pertinent for determining recovery to facilitate maximum readability. Note that system 23 is included for ease of comparison with values in Figure \ref{fig:timeseries_rvs}. The CSV file containing the full table can be found at \url{https://github.com/mtagliavia/critical-RV-experiment}.

\begin{deluxetable}{lcl}
    \vspace*{-2cm}
    \tablecaption{Description of Table/CSV Parameters \label{tab:param_defs}}
    \tablehead{
    \colhead{Parameter} &
    \colhead{Unit} &
    \colhead{Description}
    }
    \startdata
    \texttt{(index)} & unitless & Reference number for the system; first system is System 1. \\
    \texttt{period\_days\_inj} & days & Orbital period (in days) of the planet injected into the system. \\
    \texttt{period\_years\_inj} & years & Orbital period (in years) of the planet injected into the system, calculated after generation. \\
    \texttt{mass\_inj} & $M_J$ & Mass of the planet injected into the system. \\
    \texttt{tau\_inj} & unitless & Linear phase shift of the planet in its orbit, i.e. $1-\tau$ is the orbit fraction completed by a planet at the \\
    {} & {} & time of first RV measurement. \\ 
    \texttt{HIRES\_offset\_inj} & m/s & Offset from zero for the \texttt{HIRES\_synth} data. \\
    \texttt{KPF\_offset\_inj} & m/s & Offset from zero for the \texttt{KPF\_synth} data. \\
    \texttt{HIRES\_jitter\_inj} & m/s & RV jitter for the \texttt{HIRES\_synth} data, i.e. this value is the standard deviation used to generate this \\
    {} & {} & Gaussian-distributed error. \\
    \texttt{KPF\_jitter\_inj} & m/s & RV jitter for the \texttt{KPF\_synth} data, i.e. this value is the standard deviation used to generate this \\
    {} & {} & Gaussian-distributed error. \\
    \texttt{HIRES\_intvar\_inj} & m/s & Intrinsic instrument RV variability for the \texttt{HIRES\_synth} data, i.e. this value is the standard deviation used \\
    {} & {} & to generate this Gaussian-distributed error. \\
    \texttt{KPF\_intvar\_inj} & m/s & Intrinsic instrument RV variability for the \texttt{KPF\_synth} data, i.e. this value is the standard deviation used \\
    {} & {} & to generate this Gaussian-distributed error. \\
    \texttt{period\_fit\_wo} & days & Orbital period (in days) within the best fit to the data without the critical RV, as recovered by Octofitter. \\
    \texttt{period\_err\_wo} & days & Standard deviation of the associated period posterior. \\
    \texttt{mass\_fit\_wo} & $M_J$ & Mass within the best fit to the data without the critical RV, as recovered by Octofitter. \\
    \texttt{mass\_err\_wo} & $M_J$ & Standard deviation of the associated mass posterior. \\
    \texttt{ecc\_fit\_wo} & unitless & Orbital eccentricity within the best fit to the data without the critical RV, as recovered by Octofitter. \\
    \texttt{ecc\_err\_wo} & unitless & Standard deviation of the associated eccentricity posterior. \\
    \texttt{tau\_fit\_wo} & unitless & Best fit value of the orbit fraction completed by a planet after periastron at the time of first RV \\
    {} & {} & measurement, as recovered by Octofitter from the data without the critical RV. \\
    \texttt{tau\_err\_wo} & unitless & Standard deviation of the associated tau posterior. \\
    \texttt{HIRES\_offset\_fit\_wo} & m/s & \texttt{HIRES\_synth} offset within the best fit to the data without the critical RV, as recovered by Octofitter. \\
    \texttt{HIRES\_offset\_err\_wo} & m/s & Standard deviation of the associated \texttt{HIRES\_synth} offset posterior. \\
    \texttt{KPF\_offset\_fit\_wo} & m/s & \texttt{KPF\_synth} offset within the best fit to the data without the critical RV, as recovered by Octofitter. \\
    \texttt{KPF\_offset\_err\_wo} & m/s & Standard deviation of the associated \texttt{KPF\_synth} offset posterior. \\
    \texttt{HIRES\_jitter\_fit\_wo} & m/s & \texttt{HIRES\_synth} jitter within the best fit to the data without the critical RV, as recovered by Octofitter. \\
    \texttt{HIRES\_jitter\_err\_wo} & m/s & Standard deviation of the associated \texttt{HIRES\_synth} jitter posterior. \\
    \texttt{KPF\_jitter\_fit\_wo} & m/s & \texttt{KPF\_synth} jitter within the best fit to the data without the critical RV, as recovered by Octofitter. \\
    \texttt{KPF\_jitter\_err\_wo} & m/s & Standard deviation of the associated \texttt{KPF\_synth} jitter posterior. \\
    \texttt{logpost\_wo} & unitless & Log-likelihood of the best fit to the data without the critical RV, as recovered by Octofitter. \\
    \texttt{rec\_wo} & unitless & Boolean that is \texttt{True} if the recovered parameters meet the criteria from Subsection \ref{subsec:rec_criteria}, \texttt{False} otherwise. \\
    \texttt{period\_fit\_w} & days & Orbital period (in days) within the best fit to the data with the critical RV, as recovered by Octofitter. \\
    \texttt{period\_err\_w} & days & Standard deviation of the associated period posterior. \\
    \texttt{mass\_fit\_w} & $M_J$ & Mass within the best fit to the data with the critical RV, as recovered by Octofitter. \\
    \texttt{mass\_err\_w} & $M_J$ & Standard deviation of the associated mass posterior. \\
    \texttt{ecc\_fit\_w} & unitless & Orbital eccentricity within the best fit to the data with the critical RV, as recovered by Octofitter. \\
    \texttt{ecc\_err\_w} & unitless & Standard deviation of the associated eccentricity posterior. \\
    \texttt{tau\_fit\_w} & unitless & Best fit value of the orbit fraction completed by a planet after periastron at the time of first RV \\
    {} & {} & measurement, as recovered by Octofitter from the data with the critical RV. \\
    \texttt{tau\_err\_w} & unitless & Standard deviation of the associated tau posterior. \\
    \texttt{HIRES\_offset\_fit\_w} & m/s & \texttt{HIRES\_synth} offset within the best fit to the data with the critical RV, as recovered by Octofitter. \\
    \texttt{HIRES\_offset\_err\_w} & m/s & Standard deviation of the associated \texttt{HIRES\_synth} offset posterior. \\
    \texttt{KPF\_offset\_fit\_w} & m/s & \texttt{KPF\_synth} offset within the best fit to the data with the critical RV, as recovered by Octofitter. \\
    \texttt{KPF\_offset\_err\_w} & m/s & Standard deviation of the associated \texttt{KPF\_synth} offset posterior. \\
    \texttt{HIRES\_jitter\_fit\_w} & m/s & \texttt{HIRES\_synth} jitter within the best fit to the data with the critical RV, as recovered by Octofitter. \\
    \texttt{HIRES\_jitter\_err\_w} & m/s & Standard deviation of the associated \texttt{HIRES\_synth} jitter posterior. \\
    \texttt{KPF\_jitter\_fit\_w} & m/s & \texttt{KPF\_synth} jitter within the best fit to the data with the critical RV, as recovered by Octofitter. \\
    \texttt{KPF\_jitter\_err\_w} & m/s & Standard deviation of the associated \texttt{KPF\_synth} jitter posterior. \\
    \texttt{logpost\_w} & unitless & Log-likelihood of the best fit to the data with the critical RV, as recovered by Octofitter. \\
    \texttt{rec\_w} & unitless & Boolean that is \texttt{True} if the parameters recovered meet the criteria from Subsection \ref{subsec:rec_criteria}, \texttt{False} otherwise. \\
    \enddata
    \tablecomments{Due to periastron being undefined for circular orbits, the injected \texttt{tau\_inj} vs. the recovered \texttt{tau\_fit\_wo} and \texttt{tau\_fit\_w} quantify different linear phases within the planet's orbit that are only reconcilable when the recovered longitude of periastron is also considered.}
\end{deluxetable}

\begin{deluxetable*}{ccccccccccccc}
\rotate
\tablewidth{0pt}
\tablecaption{Injected and Recovered Values, Abridged\label{tab:inj_rec_values}}
\tablehead{
\colhead{\texttt{(index)}} &
\colhead{\texttt{period\_days\_inj}} &
\colhead{\texttt{mass\_inj}} &
\colhead{\texttt{period\_fit\_wo}} &
\colhead{\texttt{period\_err\_wo}} &
\colhead{\texttt{mass\_fit\_wo}} &
\colhead{\texttt{mass\_err\_wo}} &
\colhead{\texttt{ecc\_fit\_wo}} &
\colhead{\texttt{period\_fit\_w}} &
\colhead{\texttt{period\_err\_w}} &
\colhead{\texttt{mass\_fit\_w}} &
\colhead{\texttt{mass\_err\_w}} &
\colhead{\texttt{ecc\_fit\_w}} \\
}
\startdata
    1 & 6.35e+03 & 1.47 & 6.29e+03 & 1.96e+05 & 3.03 & 239 & 0.355 & 6.22e+03 & 272 & 3.19 & 2.00 & 0.409 \\
    2 & 5.54e+03 & 3.90 & 5.53e+03 & 79.9 & 3.91 & 0.114 & 0.172 & 5.56e+03 & 47.2 & 3.89 & 0.101 & 0.126 \\
    3 & 4.41e+03 & 11.2 & 4.4e+03 & 16.5 & 11.5 & 0.0718 & 0.0123 & 4.4e+03 & 17.0 & 11.4 & 0.0709 & 0.0122 \\
    4 & 1.65e+04 & 2.90 & 2.61e+04 & 2.27e+04 & 5.84 & 10.6 & 0.14 & 5.62e+04 & 2.04e+04 & 30.9 & 12.0 & 0.314 \\
    5 & 7.57e+03 & 5.85 & 7.76e+03 & 2.29e+03 & 7.27 & 2.47 & 0.0859 & 7.8e+03 & 247 & 6.44 & 0.750 & 0.068 \\
    $\vdots$ & $\vdots$ & $\vdots$ & $\vdots$ & $\vdots$ & $\vdots$ & $\vdots$ & $\vdots$ & $\vdots$ & $\vdots$ & $\vdots$ & $\vdots$ & $\vdots$ \\
    23 & 8.11e+03 & 4.46 & 3.91e+03 & 2.48e+03 & 1.46 & 1.85 & 0.305 & 8.21e+03 & 175 & 3.93 & 0.488 & 0.0376 \\
    $\vdots$ & $\vdots$ & $\vdots$ & $\vdots$ & $\vdots$ & $\vdots$ & $\vdots$ & $\vdots$ & $\vdots$ & $\vdots$ & $\vdots$ & $\vdots$ & $\vdots$ \\
    2000 & 1.18e+04 & 7.06 & 1.78e+04 & 1.77e+03 & 15.6 & 2.99 & 0.274 & 1.24e+04 & 597 & 7.21 & 0.264 & 0.0973 \\
\enddata
\tablecomments{This is an abridged table, with respect to both parameters and rows. For the CSV containing the full table, see \url{https://github.com/mtagliavia/critical-RV-experiment}.}
\end{deluxetable*}

\end{document}